# Micro-Navigation for Urban Bus Passengers: Using the Internet of Things to Improve the Public Transport Experience

Stefan Foell, Gerd Kortuem,
Reza Rawassizadeh
The Open University
Milton Keynes, UK
first.last@open.ac.uk

Marcus Handte, Umer Iqbal,
Pedro Marrón
University of Duisburg-Essen
Essen, Germany
first.last@uni-due.de

## ABSTRACT

Public bus services are widely deployed in cities around the world because they provide cost-effective and economic public transportation. However, from a passenger point of view urban bus systems can be complex and difficult to navigate, especially for disadvantaged users, i.e. tourists, novice users, older people, and people with impaired cognitive or physical abilities. We present *Urban Bus Navigator (UBN)*, a reality-aware urban navigation system for bus passengers with the ability to recognize and track the physical public transport infrastructure such as buses. Unlike traditional location-aware mobile transport applications, *UBN* acts as a true navigation assistant for public transport users. Insights from a six-month long trial in Madrid indicate that *UBN* removes barriers for public transport usage and has a positive impact on how people feel about public transport journeys.

## Categories and Subject Descriptors

H.4 [Information Systems Applications]: Miscellaneous

## Keywords

Bus navigation, public transport, smartphone application

## 1. INTRODUCTION

Public bus services are widely deployed in cities around the world because they provide cost-effective and economic public transportation. However, from a passenger point of view urban bus systems can be complex and difficult to navigate: London, for example has over 700 different bus routes and around 19000 bus stops, while Madrid has over 400 bus routes and 4,500 bus stops. Some cities have bus routes that change with the time of day (night buses, fast routes, etc.) and, in contrast to subways and trams, buses are susceptible to delays and temporary rerouting due to traffic congestion and construction work. These characteristics create a barrier of use that can hamper uptake of and satisfaction with bus transport, especially for tourists, novice users, older people, and people with impaired cognitive or physical abilities (we refer to them collectively as *disadvantaged passengers*).

In recent years mobile information systems have made urban transport systems and bus services much more accessible [2].

There is now a large number of online and mobile public transport journey planning tools which help passengers find the fastest or most convenient travel routes. These tools provide answers to important questions such as "Which bus do I need to take to get from A to B?", "How long will the journey take?", and "When does the next bus leave?". However, disadvantaged bus users have a much wider spectrum of information needs and require more fine-grained information to deal with what we call *micro-navigation decisions,* such as:

- "Is the bus that just arrived at the bus stop the one that I should take? "
- "Am I on the correct bus?"
- "In how many minutes do I need to get off the bus (or after how many stops do I need to get off)?"
- "Have I missed the stop I was supposed to get off at"?
- "If I am on the wrong bus or missed my stop, what do I need to do to resume my trip to the destination?"

Micro-navigation decisions are highly contextual. They depend not just on time and location but also on the user's current transport mode (standing outside of bus, riding on a bus etc.) and the concrete situation of a passenger (waiting for bus to arrive, riding on the correct bus, riding on the wrong bus, getting near the getting-off point etc.). Non-disadvantaged passengers have few problems making micro-navigation decisions as they can rely on their eye-sight, memory, prior knowledge, and general reasoning abilities. Disadvantaged users require contextual help to make effective micro-navigation decisions. In absence of such help disadvantaged passengers can experience uncertainty and anxiety which may lead to reluctance to use public buses. Existing mobile transport applications for smartphones provide extensive support for macro-navigation but almost no support for micro-navigation.

In this paper, we present *Urban Bus Navigator (UBN),* an urban navigation system for bus passengers with dedicated support for micro-navigation. The system is built upon an Internet–of–Things infrastructure connecting the passengers' mobile smartphones with Wifi-enabled buses in order to gain real-time information about the journey and transport situations of passengers. A key novelty of *UBN* is a *semantic bus ride detection* feature that identifies the concrete bus the passenger is riding on (for example, passenger is riding on Bus 1267, operating on Route 10). Inspired by the idea of navigation systems for cars [5,6], the *Urban Bus Navigator* provides continuous, just-in-time, end-to-end guidance for bus passengers (including dynamic rerouting) for all stages of a bus journey: before a user gets on the bus, during bus rides and when users get off the bus.

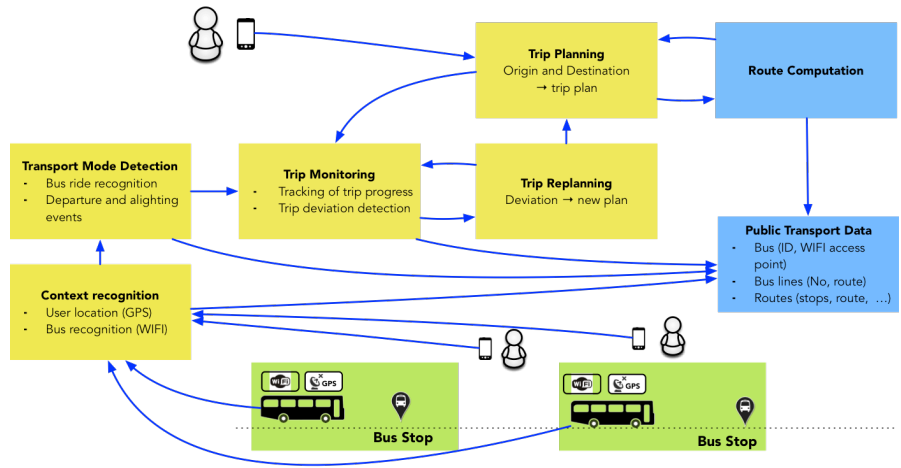

**Figure 1: Urban Bus Navigation System Architecture**

The *Urban Bus Navigator* system has been deployed and tested during a half-year long trial in Madrid. Results from technical tests indicate the real-world feasibility of semantic bus ride detection and user tests have revealed important lessons for effective user support with micro-navigation.

This paper presents the technical approach for implementing micro-navigation and reports user feedback collected during the Madrid trial.

## 2. URBAN BUS NAVIGATION SYSTEM

Most public transport journey planning tools for smartphones are context-aware and use the current time and the user's location (determined via GPS or Wifi localization) to retrieve relevant information. However time and location are coarse-grained indicators that only support macro-navigation decisions ("What is the next bus from my current location?"). Supporting micro-navigation tasks requires a fine-grained contextual, semantic understanding of the passenger's transport situations which can be achieved by enabling the passenger's smartphones to actively detect and recognize the nearby urban bus infrastructure.

*UBN* is a novel context-aware navigation system for urban bus passengers with support for macro- and micro-navigation. In the remainder of this section we will outline the UBN system architecture and describe the key innovative system functions for micro-navigation.

### 2.1 System Overview

*Urban Bus Navigator* makes urban bus travel easier for passengers by providing two innovative features not currently provided by any other public transport information system:

- Similar to a car navigation system the *Urban Bus Navigator* provides continuous dynamic navigation information for bus passengers. I.e. after the user specifies a destination and has started the navigation function, *UBN* monitors the passenger's trip progress and dynamically re-plans a trip whenever necessary.
- The *Urban Bus Navigator* supports micro-navigation (in addition to macro-navigation). I.e. *UBN* provides personalized, contextual information that address salient transport information needs, such as "Is the bus that just pulled up at the bus stop the one I should take?", "On which bus am I riding?", "Am I on the correct bus?", "How long before I need to get off?"

These user features are realized by two innovative system functions:

*Semantic Bus Ride Detection*: *UBN* detects on which bus the passenger is currently riding - not just on which bus line but on which specific vehicle. Madrid has about 2000 buses that operate daily on different routes and *UBN* is able to identify which of these buses a passenger is riding on using a Wifi-based recognition system.

*Dynamic Trip Tracking*: *UBN* uses the Semantic Bus Ride Detection (in combination with the phone's GPS) to monitor a passenger's trip progress. A passenger can either be 'on track' (i.e. the passenger is following the original plan from a location A to destination B) or 'off track' (i.e. the passenger is deviating from the plan). Trip deviations that are recognized include: a passenger is getting on the wrong bus; a passenger is riding on the wrong bus; and a passenger is getting off at a wrong bus stop. Deviations are immediately recognized and trigger re-planning of the trip which results in a new set of navigation instructions for the passenger.

### 2.2 System Architecture

The *Urban Bus Navigator* system has been developed in cooperation with Empresa Municipal de Transportes de Madrid, the Madrid transport organisation that operates the municipal bus system. Development started in early 2013 and deployment commenced in July 2013. System and user trials were conducted between October 2013 and February 2014, with new trials planned for the end of 2014. The *Urban Bus Navigator* system that is currently deployed in Madrid is composed of three key components:

- *The networked urban bus infrastructure* with Wifi-equipped buses and bus stations (Figure 1, shown in green). Wifi is used for proximity detection of buses by the passenger's mobile phone.
- *The UBN smartphone application* for bus passengers: this application provides a user interface with functions for trip planning (macro-navigation) and context-aware trip hints (micro-navigation). It also incorporates the components for context sensing, bus ride recognition, and trip tracking (Figure 1, shown in yellow).
- *Public transport data server:* this server is operated by Empresa Municipal de Transportes de Madrid and provides route planning features, real-time bus dispatching information, and estimated arrival times (Figure 1, shown in blue). We access the server via an unpublished internal web API.

In the following we provide a detailed description of design and operation of the two innovative system functions, Semantic Bus Ride Detection and Dynamic Trip Tracking.

## 2.3 Semantic Bus Ride Detection

The main task of the *UBN* bus ride detection component is to determine the current movement modality of the user, i.e. traveling in a bus or on foot, and for the former case to identify the exact bus instance the user is riding.

During an initial testing phase in Madrid, we found that relying on GPS for bus ride detection has important shortcomings in practice (high energy drain, uncertainties in urban canyons, detection of bus instance is difficult, etc.). As a practical alternative, we decided to implement our bus ride detection component by tightly integrating with the existing urban transport infrastructure. The Madrid bus system encompasses roughly 2000 vehicles that operate more than 400 routes. All buses are equipped with Wifi access points that provide free Internet access to travellers. Every access point is configured with a unique but fixed MAC address while a common SSID is used for the bus network. Thus, in order to identify bus rides, it is conceptually possible to continuously scan for the presence of a particular SSID and to identify the bus instance it is possible to use the MAC address detected during the scans. Yet, applying this approach in practice requires deeper consideration of two main challenges.

First, due to the typical Wifi transmission range of up to 100m, the access point inside the bus is also visible in a broad range around the bus. Thus, simply detecting the bus SSID as indicator for the movement modality would lead to a significant number of false classifications. While this can be mitigated by using the signal strength as an indicator for distance, we found that during bus rides the observed signal strength often ranges at the lower end of the spectrum (between -85 and -95dbm). This can be attributed to the fact that in our scenario, the Wifi access points are all mounted in the front of the bus (above the driver seat).

Second, and even more problematic, we found several cases in which the phone of a user did not detect the Wifi network of the bus for several consecutive scans, even if the user was standing in the middle of the bus. After some analysis, we found that this is a result of the ubiquity of Wifi access points in densely populated areas. In fact, in our tests we found that when moving through the city of Madrid it is commonly possible to detect more than 50 access points in a single scan, resulting in a limited number of access points that are seen due to collisions of beacon frames and limitation in the memory usage of the Wifi driver.

In order to handle these issues in an integrated manner, we developed a two-stage classifier that operates as follows. First, we determine whether the user is moving rather slowly or fast. To do this, we capture Wifi scans continuously and we analyze the change rate of the set of access points over a 30 second time frame. The assumption is that if the user is moving faster, the change rate will be higher. Using a number of test traces consisting of GPS and Wifi readings taken in Madrid, we experimentally determined a threshold for moving at a fast pace to 10 percent for cases where more than 20 access points are visible.

Using the output of this speed classifier (i.e. slow or fast), we analyze the Wifi readings as follows. If the user was not in a bus but a bus Wifi network can be seen with a very high signal strength (>-60dbm), the user is very close to the access point and thus, we consider him to be entering the bus. Alternatively, if the same bus has been seen for more than 1 minute or if the speed classifier returns driving and there is a bus Wifi available, we consider the user to be in the bus. If multiple buses are close by, we select the one with the higher signal strength. Once the user has been classified as being in the bus, we count the number of consecutive times the bus network is not received at all or at a low signal strength (<-90dbm) while the speed classifier does not signal a fast pace movement. If there are three consecutive detections with low strength at low or unknown speed, the user is considered to have left the bus.

## 2.4 Dynamic Trip Tracking

*UBN*, just like car navigation systems, monitors the user's trip progress to provide continuous contextual navigation assistance. This task is implemented by the Dynamic Trip Tracking component which implements three core tasks.

First, it formulates expectations of where a passenger is going to be next (e.g. on which bus, at which location) and what the passenger will be doing (e.g. entering or leaving a bus). This is done by building a *semantic trip plan*. Conceptually, this plan is a graph structure where nodes represents significant locations and edges represents passenger actions. Significant locations are the trip origin and trip destination, locations where a passenger enters a bus or gets off a bus. The Semantic trip plan represents multiple alternatives of how a passenger can get from the trip origin to the trip destination. A simple example represented in a semi-formal notation looks like this[1]:

```
origin | enter_bus(line_10, X, bustop_14) |
ride_on_bus (line_10, X) | get_off(bustop_45) |
destination || enter_bus(line_8, Y, bustop_14) |
ride_on_bus (line_8, y) | get_off(bustop_45) |
destination.
```

The | symbol indicates temporal sequence and || indicates a trip alternative. (This example is simplified: the actual trip plan represents each journey as a sequence of trip segments, where a trip segment represents a bus journey between two bus stops or a walking journey between two locations.)

For trip planning we use a state-of-the-art multi-modal journey planner which returns travel plans which consist of mixed walking and bus ride segments. Walking segments are described as a polyline outlining the geographic waypoints of the segment between its start and end point. In contrast, bus segments are defined by the line and direction of a bus service, as well as the stops where the user needs to get on and off the bus.

The second task of the dynamic trip tracker is to update the trip plan whenever new information is available from the Semantic Bus Ride Detector about which bus instance and trip segment the passenger is on. The trip tracker maintains a state variable which indicates the passenger's current trip segment. The transition from one segment to the next is detected by changes in semantic bus ride activity and the user's GPS position, and depends on the segment type that is investigated. A walking segment is completed once a user has arrived at the endpoint of the segment. A bus ride segment is terminated if the user has alighted at the foreseen bus stop. The completion of a segment signals successful behaviour, and the traveller can then receive information about the steps associated with the next segment.

The third and final task of the dynamic trip tracker is to detect deviations between the passenger's expected behaviour and the actual behaviour. A passenger can either be 'on track' (i.e. the passenger is following the original plan from a location A to destination B) or 'off track' (i.e. the passenger is deviating from the plan). Experience shows that deviations in public transport use are a common phenomenon, resulting from conscious decisions (passengers changing their mind halfway through a trip) or caused by accidental wrong decisions. Trip deviations that are recognized include: a passenger is getting on the wrong bus; a passenger is riding on the wrong bus; and a passenger is getting off at a wrong bus stop. If the line and direction information of the boarded bus differ from the planned bus ride, the traveller is considered to have entered the wrong bus. In contrast, if the traveller is on the correct bus but the next stop is found after the planned exit stop, the traveller is assumed to have missed his alighting. To detect this, real-time route information is retrieved for the line and direction associated with the bus taken by the user.

---

[1] The current implementation does not support trip alternatives.

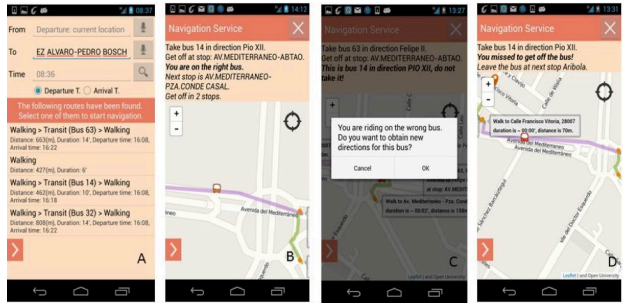

**Figure 2. Micro-and Marco-Navigation User Interface**

In case the traveller is on foot, we monitor if the planned track has been left. For this purpose, the distance between the user's current location and the current trip segment is determined. We use the minimal possible distance as an indicator for deviation and compute the shortest paths to the segment's waypoints. The length of this path is then evaluated against a spatial deviation bound. Whenever a deviation has been detected trip re-planning is triggered with the user's current position as origin.

## 3. MICRO-NAVIGATION

Bus journeys between two locations typically require walking as well as riding on a bus and consequently travellers have to perform macro-navigation tasks (e.g. finding out which bus line(s) to take to get from A to B) and micro-navigation tasks (e.g. entering the correct bus, getting off at the right bus stop).

The Urban Bus Navigator application provides dedicated support for both macro- and micro-navigation. Macro-navigation is done in the traditional way: the user can plan a route from the current location to any other location in Madrid (Figure 2a, left). In contrast, support for micro-navigation is provided via proactive information provision and requires no user intervention. Navigation information is displayed on an interface that is composed of two parts: a message area and a map area (Figure 2b-d). The map is used to help the traveller gain a geographic understanding of the entire journey, in particular to support orientation during walking segments. The current user location is shown and all travel segments involved in the journey are represented.

Assistance for micro-navigation is provided in two ways: 1) textual information is displayed in the message area above the map. This provides dynamically updated status information (which bus to get on, if passenger is on correct or incorrect bus, when to get off, etc.). 2) A subset of the textual information is graphically represented on the map as icons with variable color.

Each trip is logically composed of a sequence of trip activities and support for micro-navigation is tailored to the current activity. There are seven *trip activities*: 1) starting a journey, 2) walking to a stop, 3) arriving at a stop, 4) boarding a bus, 5) riding a bus, 6) departing a bus, and 7) walking to the destination. A concrete trip may contain any number of activities with some activities being repeated (in case of interchanges).

The micro-navigation service supports the special information needs associated with each trip activity. If the user is on foot, walking instructions are displayed and an estimation of the distance to the end location (i.e. next bus stop or destination) is given. In addition, the traveller's current location is updated on the map to support walking as a spatial navigation task. As soon as the traveller has reached the bus stop, the change of modality is prepared and information about the next bus departure is shown (bus line, direction, and exit stop). If the user is deviating from his track, a re-planning dialogue is triggered. If the user confirms this dialogue, new routes are suggested to get updated directions for the journey. If the user doesn't want to retrieve a new route at this moment, re-planning can be delayed or refused.

If a bus ride activity is recognized, the navigation system provides feedback about whether the user is on the correct bus. If the bus boarded is correct, the next stop arrivals are displayed and the user is told about the number of stops left before he needs to get off the bus. To determine this number, information about the route segment that starts at the user's current position and ends at his planned exit stop is analysed. As the bus drives along its route, the next stop arrivals are continuously updated. If only a single stop is left until the planned exit stop, the traveller is alerted about having to leave the bus soon. However, if the traveller is still found on the bus after the planned exit stop, he is supposed to have missed his alighting. In this case, the user is notified and re-planning is suggested to correct the route information. Similarly, an alert is created and re-planning is applied in case the user is found on the wrong bus.

## 4. USER EXPERIENCE EVALUATION

The *Urban Bus Navigator* has been deployed in Madrid since July 2013 and was tested during a half-year long trial. To evaluate the user experience we conducted two in-the-wild studies that involved bus passengers using the *UBN* app whenever they took a bus in Madrid.

### 4.1 Study Design

The main goal of the first study was to collect accentuated user feedback to understand the effectiveness and usability of the micro-navigation support. For this purpose, we adopted the Experience Sampling Method (ESM) and recruited 20 bus users from Madrid, 6 females and 14 males, with an age range of 32 to 55 (Mean=43, SD=7.5). In order to support ESM, a short questionnaire was embedded into the *UBN* application to collect feedback from the users "in-the-wild". The ESM component of the *UBN* application prompted users at key stages of their journey to rate their experience on a Likert scale (1 to 5) and to provide free-form textual feedback. The ESM study lasted about two months.

The goal of the second study was to gain a deeper understanding of the subjective attitudes and specific feelings the participants developed from the use of the application. To this end, we recruited another 10 participants, 3 females and 7 males, with an age range from 23 to 46 (Mean=38, SD=9.3) and conducted semi-structured interviews. With the help of a Spanish translator, we engaged the participants in a discussion about the specific situations they encountered when using *UBN*. First, the interview started with general questions about their transport habits, such as how often they conducted bus rides and for what trip purposes. Then, we asked more specific questions related to their individual experiences: How did you perceive the navigation support? Did the application change the way of how you feel about bus usage? Did the application make any difference to you? This phase lasted about one month. In the following, we report on the results and insights from both studies.

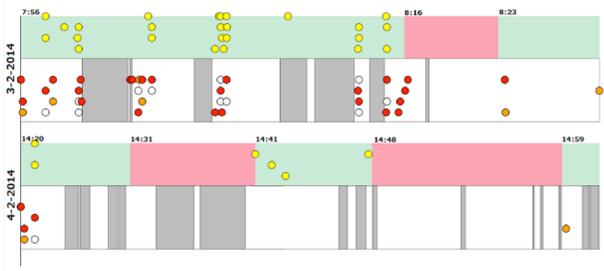

**Figure 3. Visualization of *UBN* use during two bus trips**

In addition we instrumented the *UBN* app to track application use by passengers during bus trips. We tracked use of each user-level function in the interface and captured when the Android app application was visible in the foreground and invisible in the background (see Figure 3).

## 4.2 Results

Overall we captured ESM data from 140 user trips which allowed us to identify key problem areas.

Inaccurate bus ride detection was by far the most often cited disruption to the user experience. The initial version of the bus ride detection only incorporated the presence of the bus SSID and strength of the Wifi signals. Due to the large number of buses operated in Madrid, many bus Wifi networks are constantly visible and often false positives occurred where bus rides were recognized even when the users were walking. We were able to rectify this problem and improve the user experience by implementing an improved version of the bus ride detection. The improved version additionally performed inference over the speed of movement which by far reduced the number of false positives.

A second major user concern were inconsistencies between data provided by the *UBN* app and the real-world. For example, estimated arrival times provided by the municipal transport system did at times not match actual arrival times. The second-round interviews uncovered that it was not the inconsistencies per se that worried users (users understood that estimated arrival times and other real-time data can be incorrect). However, users had higher expectations of the accuracies of data provided by *UBN* specifically because *UBN* was seemingly "smarter" – it "knew" when the user was getting on a bus and on which bus users were riding: "So if it knew that why couldn't it figure out more precisely when the bus would arrive at its stop?" We see the raised expectations as a sign that *UBN* did succeed in delivering a high-quality service to users, even though at the same time we were not able to fully satisfy the raised expectations. Furthermore we see this as evidence that the Urban Bus Navigator app is experienced as a true navigation system and used differently by passengers from traditional mobile journey planning tools.

Additional themes emerged during the interviews, primarily focused on the relationship between information availability and its impact on the participants' behaviours and feelings about bus journeys.

Similar observations were made and formulated by several participants. For instance, P1 explained: "Whenever I come to

Madrid I always need to check the signs inside the bus, and it takes time to read the moving text. This time I can check the screen on my mobile and look out the window. There is a monument, which I had read about it and I had gone past before, but never seen it. I discovered it accidentally". It became evident that bus riders often need to gather required travel information from the physical environment, e.g., the displays in buses. With our application this information became directly accessible and visible on their personal devices. This allowed them to feel more relaxed and freed time and awareness for novel travel experiences.

The discussions further revealed that public transport use is influenced by perception of its complexity that can create preferences or avoidance behaviours. For instance, P2 said: "I can't breathe properly in the subway, I prefers to use the bus, but buses are complicated ... Based on the nature of my job I always end up somewhere in the city, then I don't know how to get back home. I tried to find the nearest subway which is easy to understand ... I can do everything from beginning to end with this app and use buses. It's a combination of Google Transit and TomTom." Compared to the subway system that has only a small number of lines and routes, bus systems are often regarded as the more complex transport mode. The uncertainties involved in bus journeys can result in high barriers of use. In this regard, it is interesting that a reference to a car navigation system was made. Although it was our design intention to translate the experience of using a car navigation system to public transportation, feedback indicated that users actually did experience *UBN* as a navigation system. The fact that the application is alive during a journey appeared to increase the participants' trust in the navigation support and to raise their confidence in using the bus system.

Another insight was the surprise many participants voiced about the variety of contextual situations *UBN* could handle during their journeys and that at occasions it would reveal crucial facts that users had overlooked. As P5 explained: "Once I was waiting for my bus to ride back home ... I started killing time and playing with my smartphone. Then I got on my bus and it told me you are on the wrong bus, I checked the bus and the app was right, very smart message". In that sense, the behaviour of the application was not only perceived as information to look at, but as helpful assistance to the cognitive processes that had to be carried out. Along similar lines, P9 stated: "Putting all these things together on a single screen means it can think instead of me". In particular, the participants reported that they felt the navigation had been tailored to their personal behaviours. These feelings emerged from the fact that travel information was provided through personalized notifications that shared a relation to their trip goals, for example by alerting about the number of stops left until the next alighting.

The instrumentation of the *UBN* Android app highlighted when and how passengers were using the app during bus rides. Figure 3 shows two usage traces, i.e. it shows app usage on two separate bus journeys (by two different users). The horizontal axis indicates time from left (trip start) to right (trip end). The green areas indicate that a passenger was riding on a bus, while red indicates walking. The grey boxes in the lower half indicate when the app was in the background and thus invisible to users; the white boxes in the lower half indicates when the app was in the foreground and thus visible to users. Colored circles represent interaction events, i.e. app usage. The first trip has one interchange, i.e. the user switched buses once, while the second trip has two interchanges. Most of the app interactions shown in this diagram represent micro-navigation tasks.

Figure 3 reveals two different usage patterns. User A (top) used the app extensively during the trip (as seen by the large number of interaction events). We can see that app use is relative consistent across the journey with a slight increase in use just before the interchange. The second journey segment only has three interaction events (two just before or during on-boarding), indicating a reduced need for navigation support towards the end of the journey. User B (bottom) in contrast used the app mainly during and shortly after on-boarding, and rarely during the journey. User A extensively made use of micro-navigation features while User B did not. This is consistent with varying level of familiarity of participants with the Madrid bus system and with specific bus routes. Overall usage traces revealed a good uptake of micro-navigation features by users (yet we were not able to map app usage during journeys against user's familiarity of the bus system).

## 5. RELATED WORK

The relevance of information system for the improvement of user satisfaction with public transport system is widely recognized. As Camacho et al. anticipate [1], the main driver of future innovation in public transport are passenger-centric information services that promise novel added values for users of public transport services.

The wide-spread adoption of smartphones has stimulated a breadth of research into mobile applications to support easier public transport use. One of the first examples of a mobile transport application that has made estimated arrival times of buses accessible on mobile devices [2]. Results from a survey with a large number of application users could provide evidence on increased ridership and better satisfaction. Path2Go is a multi-modal transport application which offers real-time information about transit routes, traffic and parking [3]. The application uses GPS information for activity detection, but in contrast to our approach lacks semantic bus ride detection for identifying the bus the passenger is riding on. The same is true for commercially very successful applications like Citymapper [9]. Another recent development is Tiramisu which is able to crowd-source travel information from GPS traces of participating travellers [7]. While Tiramisu can provide real-time arrival time predictions, the system lacks specific fine-grained navigation features.

The development of navigation systems has meant a great success story, in particular for pedestrians and car drivers. Nowadays, car navigation systems are standard equipment of modern cars. Mapping the car's GPS position to a road network, drivers can be safely guided to unfamiliar location or other points-of-interests, e.g., gas stations, hotels or restaurants [6]. For pedestrian navigation, specific design challenges arise to support way finding without much distraction. For guiding pedestrians in the general direction of their destinations, Robinson et al. propose to incorporate haptic feedback on a mobile device [5]. Rehrl et al. discuss a pedestrian navigation system specifically designed for public transport buildings [4]. Their system can provide indoor navigation in interchange facilities such as train stations, using an indoor location model for way finding and Bluetooth technology for fine-grained indoor positioning.

In the current state-of-the-art, the equivalent of a navigation system for public transport use is missing. Prevailing public transport information system are mainly focused on journey planning, allowing users to retrieve a description of multi-modal route information [8]. The travellers' preferences are incorporated into constraints (e.g. time of travel) and route optimization criteria (e.g. walking distance). To relieve the user from manual input, the origin of the journey can be often automatically set to the user's current location. However, a navigation service which accompanies a traveller and decides how to best support a user from start to end of his journey based on his current activity and possible deviations is beyond the idea of classical journey planning.

## 6. CONCLUSION

In this paper, we have presented the *Urban Bus Navigator*, a navigation system for bus passengers that provides support for macro- and micro-navigation to deliver end-to-end assistance over the complete duration of a bus journey in different transport situations. *UBN* continuously tracks the behaviour of bus users to proactively display fine-grained, personalized, contextual navigation information such as whether a user has boarded the correct bus or how many stops are left until the planned exit stop. A several month-long in-the-wild study with bus users in Madrid highlighted that *UBN* is indeed experienced by passengers as true navigation system and conceived differently from existing mobile transport apps. Several positive experiences were reported by the study participants: reduced uncertainties and more relaxed travelling, better visibility and accessibility of travel information, and effective support for cognitive tasks required for bus journeys.

The design of *UBN* is generic so that it can be adapted to any city that has a digital urban transport infrastructure similar to Madrid. All in all, *UBN* demonstrates the potential of the Internet of Things for delivering innovative urban transport experiences.

ACKNOWLEDGEMENT: This work was sponsored by the EC FP7 project GAMBAS grant FP7-ICT-2011.1.3-287661